# Neural correlates of flow using auditory evoked potential suppression

Kyongsik Yun[1,2,3]*, Saeran Doh[4]*, Elisa Carrus[5], Daw-An Wu[1,2], Shinsuke Shimojo[1,2,6]

[1]Computation and Neural Systems, California Institute of Technology, Pasadena, CA 91125, USA

[2]Division of Biology, California Institute of Technology, Pasadena, CA 91125, USA

[3]BBB Technologies Inc., Seoul, South Korea

[4]Department of Food Business Management, School of Food, Agricultural and Environmental Sciences, Miyagi University, Miyagi, Japan

[5]Division of Psychology, School of Applied Sciences, London South Bank University, London UK

[6]Japan Science and Technology Agency, Saitama, Japan

*These authors contributed equally to this work.

Correspondence should be addressed to K.Y. yunks@caltech.edu

**Abstract**

"Flow" is a hyper-engaged state of consciousness most commonly described in athletics, popularly termed "being in the zone." Quantitative research into flow has been hampered by the disruptive nature of gathering subjective reports. Here we show that a passive probe (suppression of Auditory Evoked Potential in EEG) that allowed our participants to remain engaged in a first-person shooting game while we continually tracked the depth of their immersion corresponded with the participants' subjective experiences, and with their objective performance levels. Comparing this time-varying record of flow against the overall EEG record, we identified neural correlates of flow in the anterior cingulate cortex and the temporal pole. These areas displayed increased beta band activity, mutual connectivity, and feedback connectivity with primary motor cortex. These results corroborate the notion that the flow state is an objective and quantifiable state of consciousness, which we identify and characterize across subjective, behavioral and neural measures.

**Introduction**

"Flow" is a mental state of full immersion into an activity in which one is intensively engaged, accompanied with a feeling of extreme concentration, full control, achievement and pleasure in the activity [1]. In popular culture, this state is often described in professional sports as "being in the zone." According to Csikszentmihalyi (1990), flow represents the perfect experience of control of emotion and cognition during performance and learning. Verbal protocols he gathered from experts skilled in various activities, mainly via interview, indicate that the flow state can be experienced during online gaming as well as in various other activities, such as sports, education, singing, dancing, climbing, and even surgery.

Flow research has received wide interest in the field of marketing, mainly through the study of Computer Mediated Environments (CME), as a process of optimal experience during internet use[2]. It has been found that flow can create compelling consumer online experiences and can be fun and pleasurable; however, besides marketing research on flow, very little is known with regards to quantifiable measures of this experience, such as the neural dynamics underpinning this state.

Investigating the neural correlates of flow requires that the timecourse of neural data be matched to information about the states of flow during that time period. One challenge to gathering this information is that flow requires extended, uninterrupted engagement in an activity. Thus, investigations generally use post-experiment questionnaires to gather data that accompany flow experiences, this means that flow data lack timing information—they are usually blanket ratings of the entire performance period.

Several previous studies have gathered neural data from participants experiencing flow states while performing tasks [3,4], but they all possess this same limitation. Because they do not track the dynamics of flow in a way that can be aligned to the neural data, they are limited to contrasting data from different blocks of play, in which they have manipulated task conditions to either increase or decrease the likelihood of entering the flow state. These studies have provided many candidate regions and neural signatures for the flow state, but the methodology cannot dissociate between internal flow states and the external conditions that they set up to induce flow.

A study by Klasen et al. (2011) created a timecourse to align with the neural data by recording the participants' gameplay. They analyzed the game videos for events and content that would support or inhibit flow, and aligned them with the fMRI data. By finding brain areas that correlated with each content class and then calculating their overlap, they identified a cerebellar-somatosensory cortex network as a likely substrate for flow. The use of fine-scale timecourse in this study was a clear advantage, but here too the analysis actually correlated the brain activity to external game conditions rather than to data about internal states. Thus it is unclear whether this somato-motor brain network reflects the higher levels of action in the content-defined gameplay epochs, or the (inferred) higher frequency of flow experience during those epochs.

Here, we isolate the internal phenomenon of flow by correlating neural data directly with timecourses of internal states. These timecourses are based on both subjective and objective measures. Participants provided subjective data via a guided parametric version of the retropective Think Aloud design [5]. While they reviewed videos of their gameplay, they rated their experience of flow in each time segment. Objective data regarding internal states was obtained by a novel EEG probing method, which we validate and then apply here.

The EEG-based probe design is based on "telepresence" as a signature feature of the flow experience during online gaming. This refers to the gameplayer's tendency to feel like they "exist" in the game world, which leads to a neglect of sensory stimulation from the real world. The player becomes more sensitive to game-relevant sensory stimuli in the virtual word, and less sensitive to game-irrelevant, real-world stimuli [6]. Accordingly, we hypothesize that when subjects are in the flow state, they will tend to neglect sounds that aren't strictly relevant to the game. Thus, we measure the AEPs (Auditory Evoked Potentials) elicited by game-irrelevant, random beeping sounds[7] inserted throughout gameplay. Attention-based modulation of AEP amplitude is an extremely robust, and well-established effect in the evoked-related potential literature[8]. For example, a previous study showed that the AEPs elicited by a faint clicking sound were prominent when subjects counted them, but then became suppressed when subjects read a book intently[9]. Thus, we expect the AEP to be suppressed during immersion in the flow state. A recent study applied an auditory oddball as a secondary task to objectively quantify flow while playing video games [10]. Our AEP suppression has advantages in that it is completely passive and does not explicitly interfere with the subjects' cognitive processes during the flow experience.

It must be noted, however, that the AEP suppression alone is not sufficient to define the flow state during game play, because it directly captures only some of the aspects of flow, such as intense, focused concentration, and perhaps telepresence indirectly. Subjective behavioral reporting on the experience of flow is also neither sufficient nor reliable in defining flow. It was for this reason that we decided to define the flow state by a correlation of objective and subjective measures, assessing flow state by comparing the AEP amplitude suppression with the respective post-gameplay ratings of flow for each participant.

The present study aims: 1) to validate AEP suppression as a measure of the flow state, thereby providing a passive probe that can be used to periodically measure the participant's flow state without disrupting their engagement in the task. 2) to use the measured timecourse of flow to localize its neural correlates, and to characterize functional connectivity between those brain regions. By doing this, we would like to characterize the dynamics of flow in a quantitative manner, and link it to objective neural substrates.

**Methods**

**Participants**

Twenty-nine healthy subjects participated in this study (24 males, 5 females, age: 23.5±3.4 years (mean±SD)). We recruited subjects on the basis of previous experience with video games in general (average 13.6±10.1 hours/week, range 7~49 hours/week) with ads posted on the main campus of the California Institute of Technology. All participants provided written informed consent after receiving a detailed explanation of the experimental procedures. The Institutional Review Board of California Institute of Technology approved all experimental procedures and this study was carried out in accordance with the approved guidelines. Participants were excluded if they had a history of neurological disorder such as seizure, stroke, head injury, or a substance use disorder other than caffeine or nicotine. We obtained a set of self-reported flow questionnaires [11] at the end of the experiment.

**Task**

The participants played a FPS video game (Call of Duty : Modern Warfare 2, Activision). During pre-experiment game training, we adjusted the difficulty of the game depending on the skill of each participant so that it was right above the player's skills. During the experimental session, participants played the game for an hour, consisting of a 30min low challenge and a 30min high challenge game scenario (Figure 1). During the game play, randomly distributed beep sounds (inter-beep interval=1~120sec: duration=200ms; 40dB) were presented via speakers to the participants, in order to measure auditory evoked potentials.

**Behavioral analysis**

While participants were playing, we recorded the game play using Fraps, a realtime video capture software (www.fraps.com). This recording was used in the review session following game play. Here, participants were asked to review their own game play by replaying a video of their own game session and responding whether they experienced flow or not for each 5 min time period of the video. The 5 min time period was chosen based on the previous studies of neurofeedback and subjective behavioral ratings [12,13]. Player performance (i.e., number of kills – deaths) was tracked across the same time bins of game play. Flow experience and performance distribution histograms were compared using Pearson's correlation. For additional skill-based analysis, fifteen highest performing participants were grouped as "experts" and the others as "beginners."

**Insert figure 1.**

**Electrophysiological recording and analysis**

We recorded EEG activity from 128 scalp electrodes (EGI System 200; Electrical Geodesics, Eugene, OR.) during gameplay. Electrode impedance was kept under 40 kΩ for all recordings[14]. Electrode nets were covered with a shower cap to prevent electrodes from drying so that we could maintain low electrode impedance. Vertical and horizontal ocular movements were also recorded. The EEG was continuously recorded at a 500 Hz sampling frequency and filtered (high pass 0.1Hz, low pass 200Hz, notch filter 60Hz and 120Hz). The EEGs were segmented from −500ms to 1000ms relative to the onset of each beep. Ocular artifact reduction was performed using ICA component rejection in EEGLAB [15].

Following pre-processing, we computed the evoked activity aligned to each task-irrelevant beep sound, and this was done to evaluate epochs with and without AEP suppression. Consequently, we separated each epoch as flow and non-flow groups based on the participants' ratings as well as AEP suppression (Table 1) and computed each epoch's source localization and partial directed coherence values to define the network of regions involved and effective connectivity within it. The time window used for source localization and partial directed coherence was 1000ms following the beep.

**Time-frequency analysis**

The epoched data were analyzed by means of a event-related spectral perturbation (ERSP) [16] (window length, 250 ms; step, 25 ms; window overlap, 90%). This was done to select the epochs that showed AEP suppression. We removed the baseline of 500 ms preceding the beep onset with duration of 500 ms from time frequency ERSP charts. We used a nonparametric permutation test with 5000 randomizations for comparisons between the activation and baseline, corrected for multiple comparisons [17]. The corrected threshold was set to $p < 0.05$, and the time-frequency representation only shows statistically significant results.

**sLORETA**

Standardized low resolution brain electromagnetic tomography (sLORETA) [18] was used for source localization. Thousands of synchronized postsynaptic potentials from pyramidal neurons of the cortex produce scalp EEG activity [19]. sLORETA computes the three dimensional localization of these activities. The subject-specific 3D coordinates of the 128 electrode positions were estimated (Geodesic Photogrammetry System; Electrical Geodesics, Eugene, OR.) and applied to a digitized MRI version of the Talairach Atlas (McConnell Brain

Imaging Centre, Montréal Neurological Institute, McGill University). These Talairach coordinates were then used to compute the sLORETA transformation matrix for each participant. Following the transformation to an average reference, the EEG activity of the flow and non-flow epochs selected from the above time frequency analysis was used to calculate cross spectra in sLORETA for each participant. We used delta (1~4Hz), theta (4~8Hz), alpha (8~12Hz), and beta (12~30Hz) frequency bands for the following analyses. Using the sLORETA transformation matrix, the cross spectra of each participant and frequency band were then transformed into sLORETA files. These files included the 3D cortical distribution of the electrical neuronal generators for each participant. The computed sLORETA image displayed the cortical neuronal oscillators in 6239 voxels, with a spatial resolution of 5 mm [20]. We used a nonparametric permutation test with 5000 randomizations for comparisons between the flow and non-flow[17]. The threshold was set to $p < 0.01$.

**Partial directed coherence**

Partial directed coherence (PDC) is a form of frequency-domain Granger-causality, which quantifies the direction of information transfer between brain regions [21,22]. The EEG data from ROIs defined by source localization were windowed in 500-sample long intervals (i.e., 1000ms in length). The PDC values were evaluated in the delta (1~4Hz), theta (4~8Hz), alpha (8~12Hz), and beta (12~30Hz) ranges. Model order (i.e., time delay of the autoregressive parameters) was set to 15, based on previous studies for sufficient frequency resolution [23,24]. We used the nonparametric bootstrap approach for further statistical analyses [25,26]. For details of this method, see Snijders and Borgatti [26]. The threshold was set to $p < 0.001$, considering Bonferroni correction for multiple comparisions (3 ROIs, 3X3=9 comparisions). Each arrow pointing from the region $i$ to its target region $j$ represents a significant PDC. SPSS (Windows version 15.0; SPSS, Inc., Chicago, IL) was used for the statistical analyses.

**Results**

Mean duration of flow was 8.31±3.61 min (13.9% of total time). 53.0% of the flow experience occurred between the 25 and 45 minute marks of the total one-hour game session. A significant positive correlation was found between the overall occurrence of the flow experience and the performance distribution throughout the game play (Pearson's correlation, R=0.68, p=0.014) (Figure 2). These results indicate the close relationship between subjective flow experience and performance, suggesting that the higher the experience of flow, the better the gaming performance.

**Insert Figure 2.**

Next we tested AEPs as a neural marker for the flow state. Since we did not observe the conventional clear AEP waveform, due to insufficient number of trials and background noise from the game play, we analysed the ERSPs of the signal at low frequencies (1-30Hz). In EEG analysis, the classical ERP model is limited to study complex brain dynamics because the trial-by-trial frequency components tend to be averaged out by this method. ERSP has been adapted for this reason[27-30]. Given the low signal-to-noise ratio, we used ERSPs, which represent the power of oscillations of the ERP. This was done because evoked potentials predominantly contain low-frequency information, and by doing this we would still be able to evaluate the suppression of the auditory evoked potential in the absence of a clear AEP. Following established protocols, ERSPs were calculated for each tone at the averaged channels Fz and Pz[8,31]. We compared the EEG ERSP of the post-stimulus window to the baseline (500~0ms before the beep onset) and separated epochs with significant post-stimulus activation from those showing deactivation at averaged. Out of the epochs showing suppressed signal, we categorised them as flow segments only if they were also rated as flow by participants. The self-reported non-flow trials showed larger evoked potential activation than the flow trials (t(1861)=3.84, p=0.0001) (Figure 3 and Figure S1). We also found a significant correlation between suppressed evoked potential and self-reported experience of flow (Table S1. Chi-square test; $X^2(1)=97.0$, p<0.001).

**Insert Figure 3.**

Following the analysis of the auditory evoked potential, we source localized the epochs corresponding to the flow state and those corresponding to the non-flow state at time window of 0~1000ms after the stimulus. This was done to estimate the possible sources of the activity that gives rise to flow experience, and therefore to find the brain areas associated with flow. Figure 4 depicts localization results, which showed that the anterior cingulate cortex (ACC: MNI x=-6, y=25, z=21, Figure 4a); and temporal pole (TP: x=-55, y=10, z=-25, Figure 4b) were significantly activated in the flow state as compared to the non-flow state only in the beta frequency range. We also found significant beta frequency oscillations in the primary motor cortex (precentral gyrus (M1); x=15, y=-20, z=70; Figure 4c) in the beginners compared to experts.

**Insert Figure 4.**

To examine the link between these three brain regions and the experience of flow, we computed correlations between the subjective behavioral flow ratings and beta activity . We found significant positive correlations between the flow ratings and activation in ACC (Pearson's correlation; $R^2=0.22$, $P=0.017$) and TP ($R^2=0.26$, $P=0.009$) (Figure 5A and B) and a significant negative correlation with activation in M1 ($R^2=0.32$, $P=0.003$) (Figure 5C). These correlation results therefore suggest that ACC and TP are the core areas associated with the flow state.

**Insert Figure 5.**

To further understand the network properties of the neural correlates of flow, we analyzed effective connectivity using PDC across the same three regions of interest. We found that the PDC connectivity increased during the flow compared to non-flow state. More specifically, the top-down connectivity increased (ACC->M1, TP->M1) and bottom-up connectivity decreased (M1->TP) during the flow state (nonparametric jackknife procedure, $p < 0.001$) compared to non-flow state (Figure 6).

**Insert Figure 6.**

**Discussion**

The current study aimed to generate the mental state of "flow" in the laboratory and to quantify and to characterize it using a novel combination of objective and subjective data, namely the suppression of AEP, and video-assisted self-reports. By correlating the timecourse of flow with the EEG data, we localized the potential neural correlates of flow and characterized effective connectivity between these brain areas. Specifically, the ACC and TP emerged as important hubs associated with the experience of flow, showing increased beta frequency power and increased effective connectivity patterns during the flow state compared to non-flow.

There are several psychological criteria for flow [32], including (1) intense and focused concentration, (2) a loss of self-consciousness, and (3) a feeling of control and ease of performance. For example, when a game player experiences flow, his/her attention is completely focused on the game character and the opponent, and he/she ignores any task-irrelevant stimuli. The actions he/she performs are perceived as if they were his/her own, rather than in the virtual world. We therefore exploited this notion by evaluating modulation of the AEP amplitude in flow and non-flow. We found significantly suppressed neural activity following the beep when in flow, compared to non-flow in the oscillatory evoked activity, which represents the power of oscillations of the ERP. This result suggests that when participants experienced flow, their brains shut off task-irrelevant stimuli in the literal sense (even at the sensory level). Through the use of the AEP, we were able to quantify the flow state without disrupting the participant's engagement with the game.

Our source localization results are consistent with the flow characteristics mentioned above. Activation of the ACC known to be involved in the processing attention and focus [33,34], which is in line with the first criterion of flow. The TP, an empathy-related brain region, was also activated and we interpret this in light of the second characteristic of flow, a loss of self-consciousness and telepresence. The TP has been known to be crucial in the process of distinguishing self vs. other as well as in perspective-taking in social affective processes[35,36]. Lastly, our current results show a pattern of negative correlation between motor activity and the flow experience. We also found that the motor activity decreased in experts compared to beginners. These results are consistent with previous studies, which have found that motor activity decreases when motor behavior is processed more efficiently[37,38], and that experts develop a focused and efficient organization of neural networks[39]. This accords with the third characteristic of flow, which is a feeling of control and ease of performance.

Our effective connectivity results showed a top-down connectivity from the ACC and the TP to M1 in the flow state. Enhanced top-down processing has been known to be consistent with heightened performance and concentration[40,41]. However, our top-down connectivity results are different from the typical top-down executive control represented by explicit (conscious) cognitive process of the prefrontal cortex. Rather, ACC and TP performed as implicit (unconscious) cognitive and social processes, including extreme focus, empathy, and telepresence. In summary, we argue that the ACC and the TP are the two main regions that may provide the neural basis of the flow state, or "being in the zone".

Our study took several measures to maximize the likelihood of inducing a deep flow state. First, we used a first-person-shooter video game known to be very addictive[42], and our participants were all experienced players, many of whom routinely devoted large amounts of time to gameplay. This allowed us to recreate a state of flow in the laboratory in players who had spontaneously experienced flow in the past. The laboratory setting was comfortable, and provided a typical computer game playing environment. This contrasts with an MRI environment, where players are prone, immobilized and in a confined space. Flow is less likely to occur if the environment is unfamiliar and/or uncomfortable[43]. Most importantly, we allowed the participants to play the game continuously for an hour. According to our results, 87.5% of the participants require at least 25min to get into the flow state (Figure 2). This suggests that the relative strength of flow experiences in previous neuroscientific studies may be limited, as individual blocks of their experiments range from 3-12 minutes.

Lastly, we implemented an objective electrophysiological evaluation (AEP suppression) to test whether participants were in flow and cross-validated these results through subjective behavioral flow ratings. This technique allowed us to track the strength of flow and build a timeline of the flow experience. Previous studies have been limited in their characterization of flow: they tended to quantify objective external features such as game performance (killing vs. being killed, active fighting vs. boring scenes in the case of shooting games; easy vs. hard levels in general). Although there is a positive correlation between such external criteria and the flow experience, using them as substitutes for direct measures of flow causes clear confounds in the analysis of neural correlates. We argue that our cross-validation method (across neural, behavioral and subjective measures) captures the actual emergence of flow in the closest possible way to the original definition and the phenomenology of the state.

One limitation of our study is related to our use of EEG, in that it cannot capture the activity of deep brain structures, such as amygdala and midbrain reward circuits. These regions have been known to be closely related with empathy, attention, motivation, and reward, which are highly probable to be correlated with flow experience[44]. Future studies using different techniques are necessary to pursue such targets. For example a combination of EEG and fMRI could be used to simultaneously measure the AEPs and image deep brain areas.

We should also mention the limitation that we did not resolve a clear AEP waveform, probably because our experimental setting was ill-suited to extracting classical ERPs through trial averaging. Rather than having hundreds of regularly repeated trials, we had a relatively small number of beeps randomly inserted into the dynamic and irregular auditory game environment. However, anlaysis in the time-frequency domain revealed clear patterns of evoked oscillatory activity, showing large positive components of the AEP phase-locked to the beep onset and suppression of those components during the flow state.

Techniques of detection and quantification of the flow state can contribute not only to neuroscience of the altered state of consciousness, but also to game content design. The behavioral characteristics of the flow experience, including its mean duration, temporal location, and the relationship with gaming performance, would be very useful for effective, immersive game design as well as marketing research on internet shopping, etc. Degrees of the feeling of being in the zone have been known to be highly associated with amount of information search in game players and purchasing intention among internet shoppers[45]. Thus, whether our technique can be generalized to other situations of flow, particularly in the domain of marketing, would be an intriguing question to address in future. Thus all together, the flow is a psychological and neurophysiological phenomenon which can be uniquely defined, and feasibly assessed by a combination of subjective and objective techniques as demonstrated in the current study. Whether or not the neural correlates and the neural measure are applicable to the flow state in other kinds of context, such as sports, gambling, singing, dancing, internet surfing/shopping, etc. would be an obvious next step of future research.

**Acknowledgments**

This work was supported by JST-ERATO, JST-CREST, Tamagawa-Caltech gCOE programs, Grant-in-Aid for Scientific Research (Kakenhi) in Japan, and Basic Science Research Program through the National Research Foundation of Korea (NRF) funded by the Ministry of Education, Science and Technology (2013R1A6A3A03020772).

**Author Contributions**

K.Y., S.D., E.C., and S.S. designed and performed the experiments. K.Y. analyzed the data. K.Y., S.D., E.C., and S.S. wrote the manuscript.

**Additional information**

Competing financial interests: The authors declare no competing financial interests.

**Figure Legends**

Figure 1. Experimental procedure.

Participants played a first-person shooter video game for an hour, consisting of 30min low challenge and 30min high challenge. Randomly distributed beep sounds (inter-beep interval=1~120sec) were presented to participants via speakers during the game.

Figure 2. Histograms of (A) the flow distribution and (B) the performance distribution across time of game play. Performance was defined as the mean number of "kills" minus "deaths" across all participants (R=0.68, p=0.014).

Figure 3. Time-frequency analyses of auditory evoked activity during (A) flow and (B) non-flow states. The non-flow epochs showed significant evoked potential activation and the flow trials showed significant evoked potential suppression compared to the baseline (500~0ms before the beep onset) (average of channels Fz and Pz; nonparametric permutation test, p<0.05). The vertical line represents the onset of the beep.

Figure 4. (A) Source localization showing the contrast of flow state minus non-flow state for (A) the anterior cingulate cortex (ACC) (x=-6, y=25, z=21), and (B) temporal pole (TP) (x=-55, y=10, z=-25). (C) the flow minus non-flow contrast for beginners minus experts in the precentral gyrus (M1) (x=15, y=-20, z=70) (all: non-parametric permutation test, red: p<0.05, yellow: p<0.01).

Figure 5. Correlation between the behavioral flow ratings and source localized EEG activity. For the (A) temporal pole (TP) ($R^2$=0.26, P=0.009), (B) anterior cingulate cortex (ACC)

($R^2$=0.22, P=0.017), and (C) primary motor cortex (M1) ($R^2$=0.32, P=0.003).

Figure 6. Effective connectivity using PDC between regions of interest, including anterior cingulate cortex (ACC), temporal pole (TP), and primary motor (M1) cortex. These regions were selected based on a priori source localized activity in Figure 4. Red arrows indicate the contrast for PDC connectivity of flow with non-flow, and the blue arrows indicate the contrast of non-flow with flow.

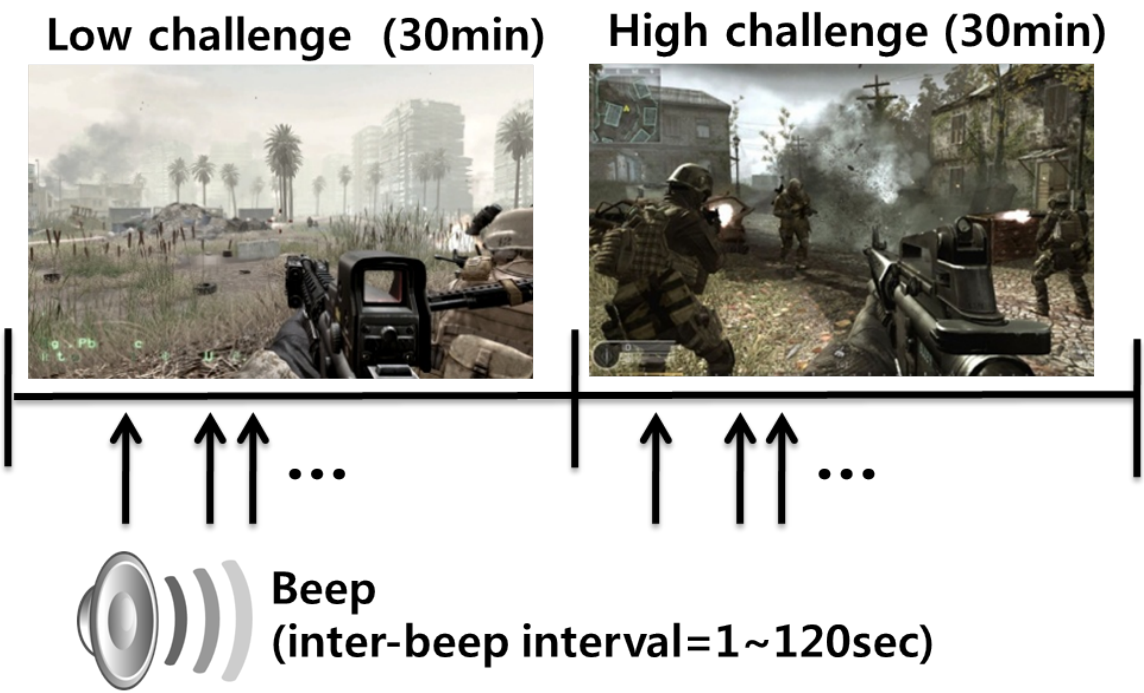

Figure 1

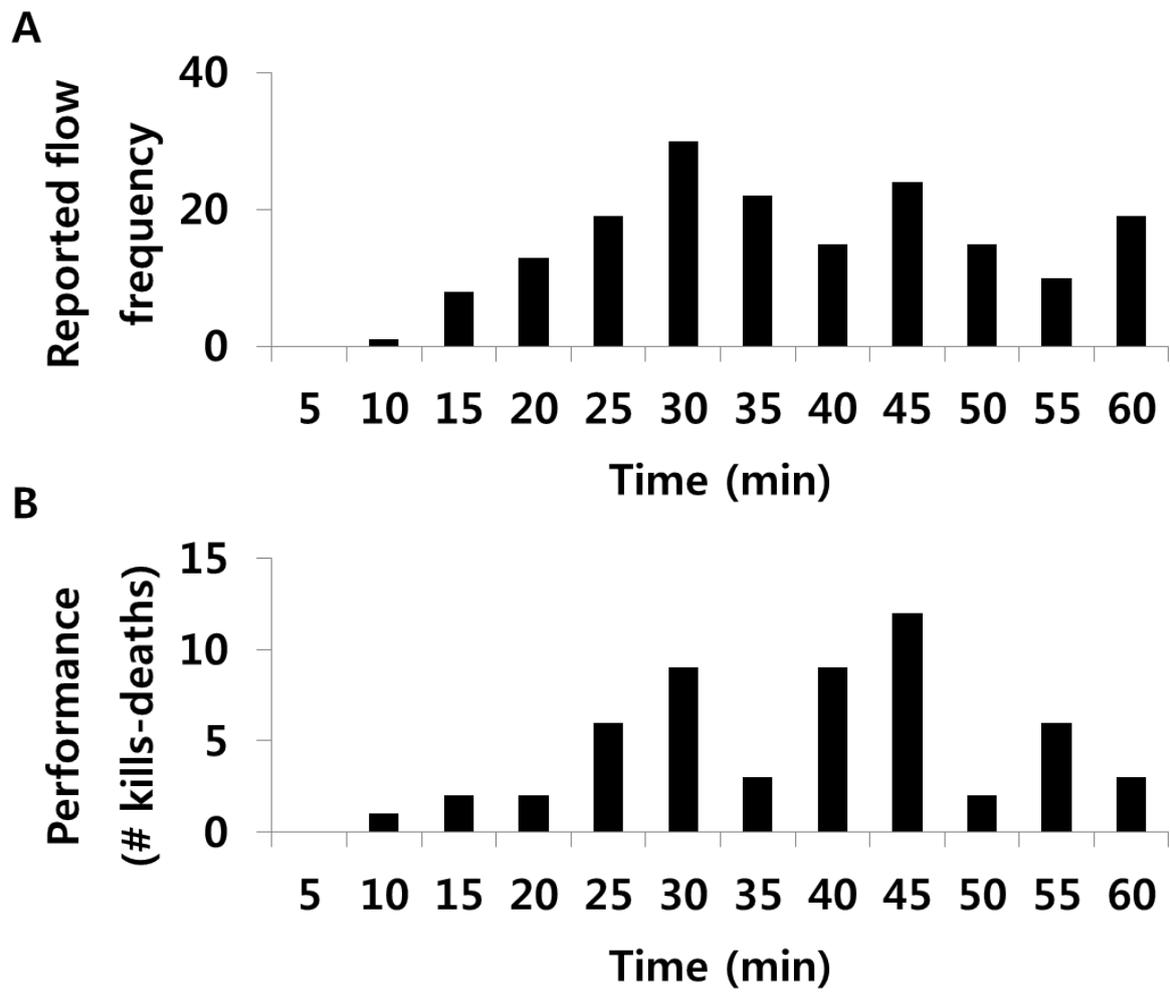

Figure 2

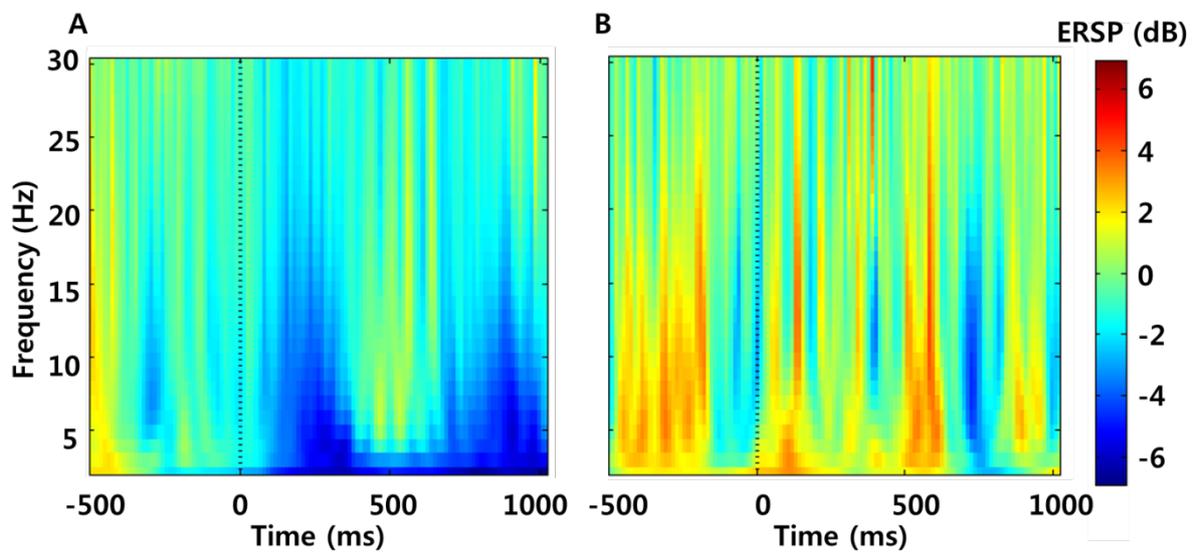

Figure 3

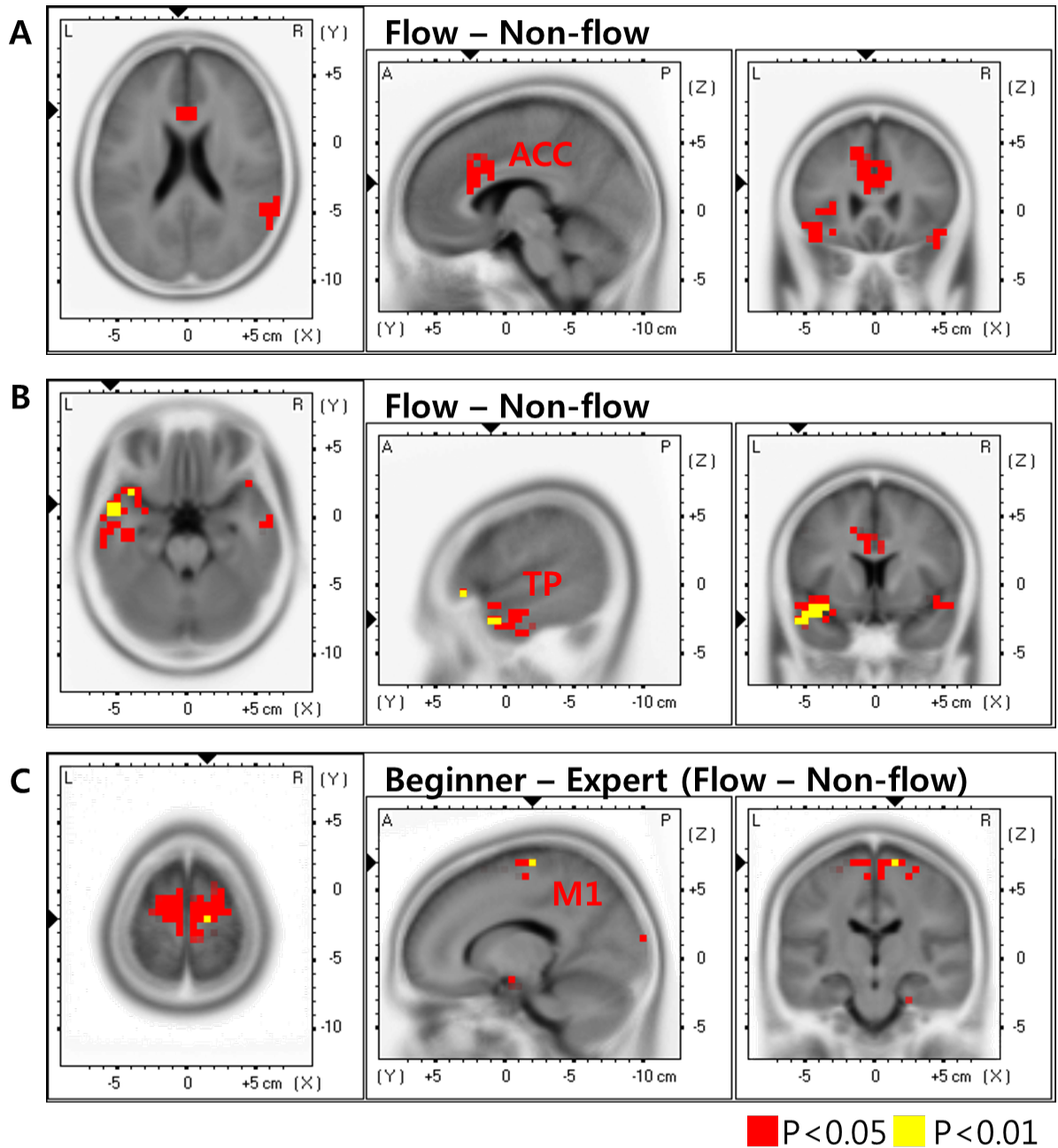

Figure 4

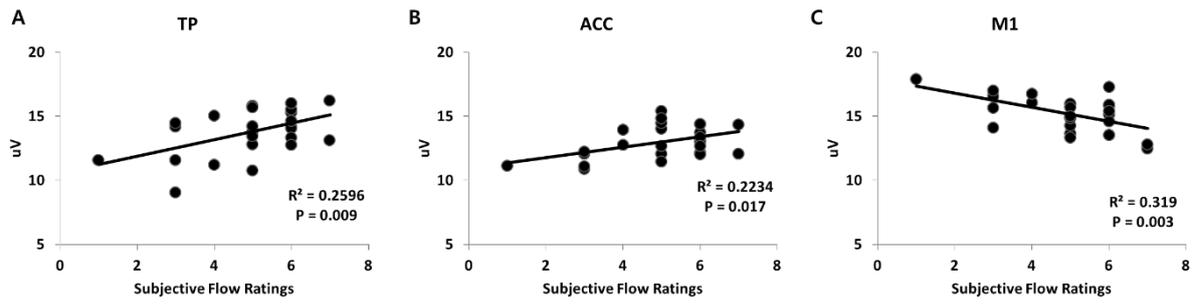

Figure 5

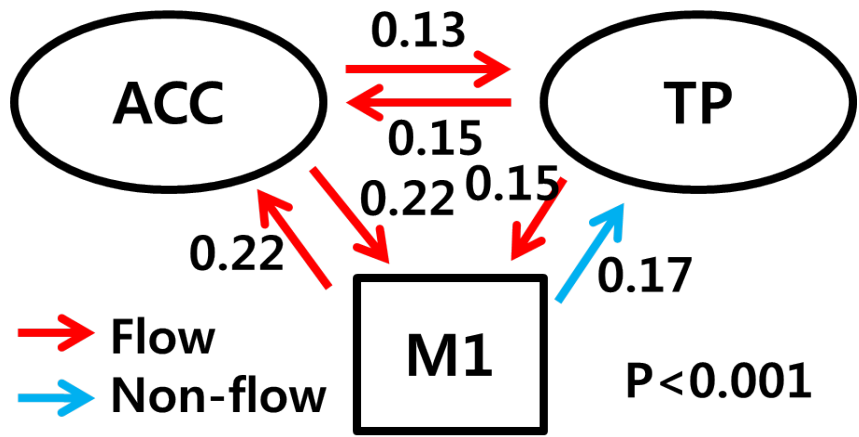

Figure 6

# Supplementary Information

# Neural correlates of flow using auditory evoked potential suppression

Kyongsik Yun[1,2,3*], Saeran Doh[4]*, Elisa Carrus[5], Daw-An Wu[1,2], Shinsuke Shimojo[1,2,6]

[1]Computation and Neural Systems, California Institute of Technology, Pasadena, CA 91125, USA

[2]Division of Biology, California Institute of Technology, Pasadena, CA 91125, USA

[3]BBB Technologies Inc., Seoul, South Korea

[4]Department of Food Business Management, School of Food, Agricultural and Environmental Sciences, Miyagi University, Miyagi, Japan

[5]Division of Psychology, School of Applied Sciences, London South Bank University, London UK

[6]Japan Science and Technology Agency, Saitama, Japan

*These authors contributed equally to this work.

Correspondence should be addressed to K.Y. yunks@caltech.edu

S1 Table. Percentage of epochs with and without suppressed AEP rated as flow and non-flow

|  | AEP suppression | No AEP suppression |
|---|---|---|
| Flow rating | 26.1% | 18.7% |
| Non-flow rating | 14.6% | 40.6% |

Epochs rated as flow showed significantly larger number of suppressed AEP compared with the AEP of epochs rated as non-flow (chi-square test; $X^2(1)=97.0$, $p<0.001$).

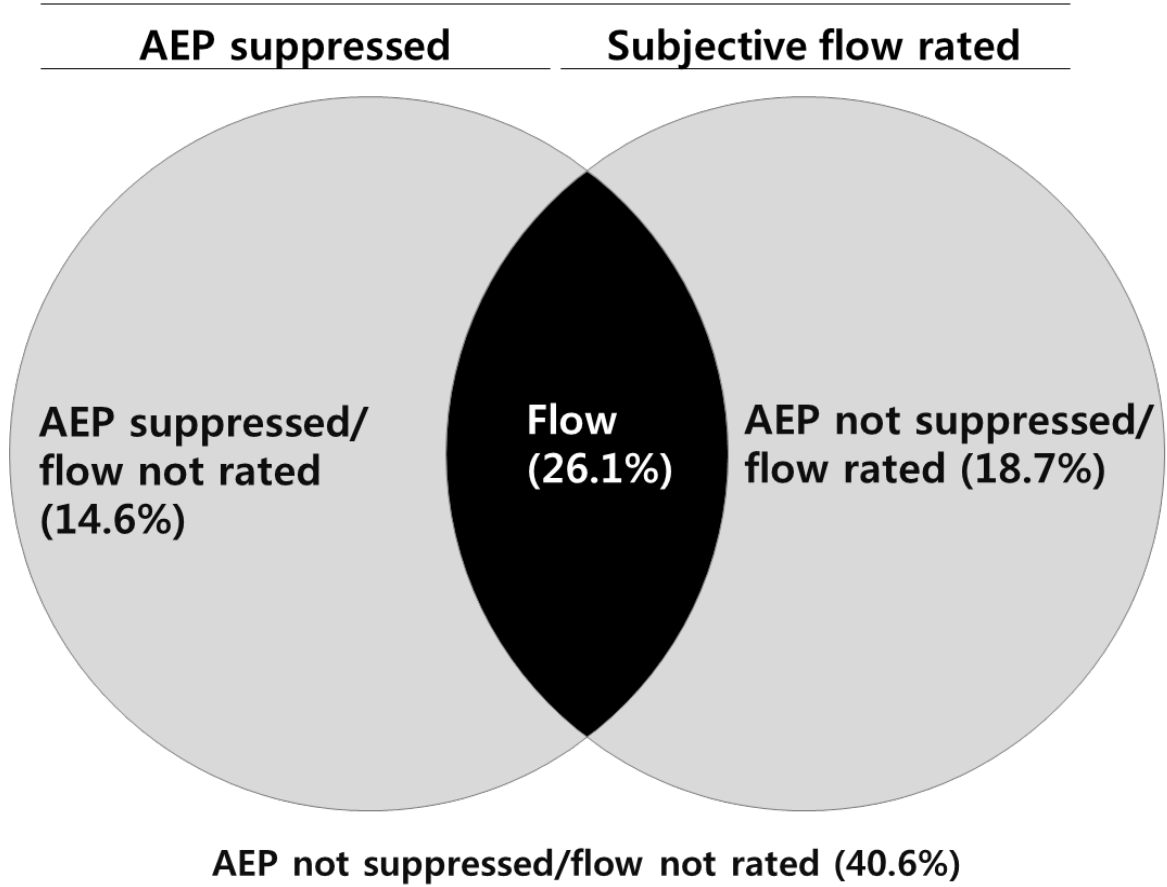

Table S1. Venn diagram of the percentage distribution of all the trials

S1 Figure. Peak auditory evoked potential (0~500ms) in non-flow and flow state. The non-flow trials showed larger evoked potential activation than the flow trials (t(1861)=3.84, p=0.0001).

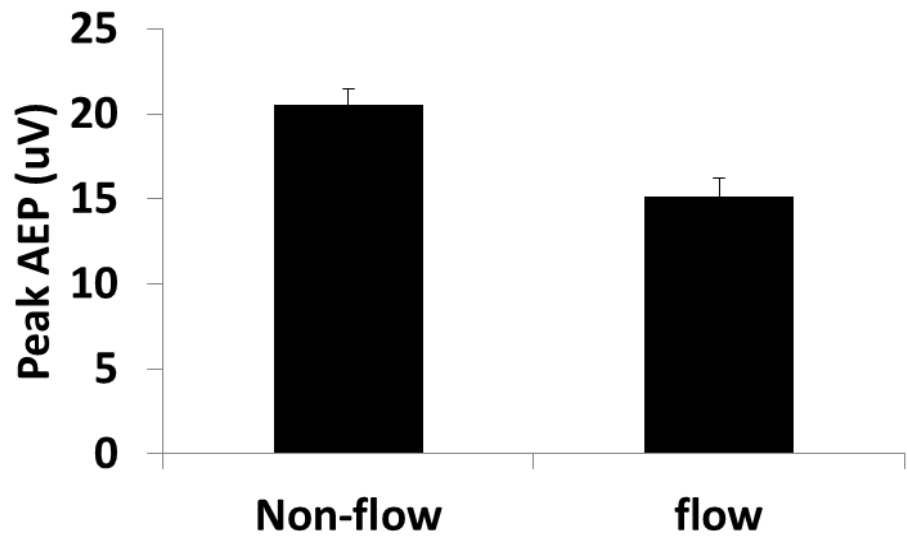